# Synthesis, Contact Printing, and Device Characterization of Ni-Catalyzed, Crystalline InAs Nanowires

Alexandra C. Ford[1,2], Johnny C. Ho[1,2], Zhiyong Fan[1,2], Onur Ergen[1], Virginia Altoe[3], Shaul Aloni[3], Haleh Razavi[1], Ali Javey[1,2,*]

1. Department of Electrical Engineering and Computer Sciences, University of California at Berkeley, Berkeley, CA 94720.
2. Materials Sciences Division, Lawrence Berkeley National Laboratory, Berkeley, CA 94720
3. *Molecular Foundry, Lawrence Berkeley National Laboratory, Berkeley, CA 94720*

* Author to whom correspondence should be addressed.
Electronic mail: ajavey@eecs.berkeley.edu

ASTRACT - InAs nanowires have been actively explored as the channel material for high performance transistors owing to their high electron mobility and ease of ohmic metal contact formation. The catalytic growth of non-epitaxial InAs nanowires, however, has often relied on the use of Au colloids which is non-CMOS compatible. Here, we demonstrate the successful synthesis of high yield of crystalline InAs nanowires with high yield and tunable diameters by using Ni nanoparticles as the catalyst material on amorphous $SiO_2$ substrates. The nanowires show superb electrical properties with field-effect electron mobility ~2,700 $cm^2/Vs$ and $I_{ON}/I_{OFF}$ >$10^3$. The uniformity and purity of the grown InAs nanowires are further demonstrated by large-scale assembly of parallel arrays of nanowires on substrates via the contact printing process that enables high performance, “printable” transistors, capable of delivering 5-10 mA ON currents (~400 nanowires).

The ability to control the size, structure, composition and morphology of semiconductor nanowires (NWs) makes them ideal one-dimensional building blocks for potential applications in high performance nanoelectronics and large-area, flexible electronics.[1-7] Uniquely, nanowires can be readily assembled on various substrates using low temperature processing conditions, therefore, making them compatible with CMOS processing while avoiding the lattice mismatch and single-crystalline growth challenges often encountered for epitaxial, planar thin films.[8-11] As a result, hybrid electronics[4] consisting of "top-down" Si CMOS and "bottom-up" nanomaterials may be envisioned for enabling advanced functionalities. In particular, InAs nanowires have been widely explored as the channel material for high performance transistors owing to (i) their high electron mobility ($\mu_n$), and (ii) the ability to readily form near-ohmic, transparent metal source/drain contacts due to their small band gap and the intrinsic surface charge accumulation layer.[2,12-15] Additionally, InAs has a large bulk exciton Bohr radius (~34 nm) which is on the order of the radial size of nanowires, resulting in 1-D quantum confinement of the carriers with potentially interesting carrier transport characteristics.[16]

The reported synthesis of non-epitaxial, semiconductor nanowires often involves the vapor-liquid-solid (VLS) or vapor-solid-solid (VSS) mechanisms, where a metal nanoparticle (NP) catalyzes the growth. Currently, InAs NWs grown on amorphous substrates have all utilized Au NPs as the catalyst material.[12-15,17] Au, however, is a non-ideal material for electronics and is incompatible with conventional Si CMOS technology because of its well-known role as a deep-level trap in Si, resulting in severe degradation of CMOS properties. This presents a challenge for potential integration of InAs NWs with Si CMOS for hybrid electronics. Furthermore, the exact magnitude of Au incorporation and its role on the electrical properties of InAs NWs is still not known. While there has been multiple reports of growing InAs NWs epitaxially on crystalline Si and InAs substrates without the use of metal NPs,[18-19] such approaches are not compatible with certain applications, for instance for those utilizing roll or contact printing of NWs that require high yield growth on amorphous substrates (i.e., glass rollers).[8-9] Therefore, there is a need to explore alternative catalyst materials for VLS/VSS growth of high purity, crystalline InAs NWs for high performance, CMOS-compatible, and yet printable electronics. To address this challenge, here, we demonstrate the successful synthesis of crystalline, high-mobility InAs nanowires by using Ni NPs with tunable diameters. These grown nanowires are then successfully transferred to receiver substrates in highly regular arrays by the contact printing process and configured as the channel material for high performance transistors.

Ni nanoclusters used for InAs nanowire growth were obtained by thermal annealing (800-900ºC) of thin Ni films (thermally evaporated) on Si/$SiO_2$ (50 nm thermally grown) substrates in a hydrogen environment. Due to the finite mobility and diffusion of the Ni atoms on $SiO_2$ surfaces at elevated temperatures, NPs are formed upon aggregation and bonding of the atoms via metal-metal interactions. The NP diameters can be readily tuned by the corresponding thin film thickness and the annealing conditions. Figure 1 shows the atomic force microcopy (AFM) images and corresponding particle diameter distributions of Ni particles formed by the thermal annealing of 0.5 nm (Fig. 1a and b), 1.5 nm (Fig. 1c and d), and 3 nm (Fig. 1e and f) Ni films at 850 ºC for 10 min. The NP diameters obtained from AFM and scanning electron microscopy (SEM) for the 0.5, 1.5, and 3 nm films are 10±2, 14±3, and 26±5 nm, respectively. The diameter variation as a percent of the mean for all three particle sizes is ~20%. This is remarkably good considering the simplicity of this method, and that the variation for commercially available colloidal Au NPs used to grow nanowires of similar diameter (20-40 nm) is ~10%.

The NPs attained from thin film annealing were used as catalytic seeds for the growth of InAs nanowires. After the thermal annealing process, the sample temperature was reduced to 470-550ºC, and InAs NWs were then grown for ~1 hr by vaporization of InAs solid source (source temperature 720ºC). The growth furnace consisted of two independently controlled temperature zones, one for the solid source and the other for the sample, similar to the previously reported Au-catalyzed InAs NW growth set up.[13] Hydrogen (150 s.c.c.m.) was used as the carrier gas for the delivery of the thermally vaporized solid InAs source. The pressure was maintained constant at ~1 torr. The NWs were grown chemically intrinsic without any intentional doping. SEM images of Ni-catalyzed InAs NWs grown from different particle diameters are shown in Fig. 2. From the SEM images, it is evident that the grown NWs are relatively straight with low structural defect density. Furthermore, Ni NPs can be clearly observed at the tip of most nanowires (Fig. 2), which is a distinct characteristic of the tip-based, VLS/VSS growth mechanism. The NW diameter shows a direct correlation with the catalytic NP diameter. Nanowire diameters of 23±6, 26±8, and 38±9 nm were obtained for 10, 14, and 26 nm NPs, respectively. The variation as a percent of the mean for the grown nanowires is 25-32% which is slightly larger than that of the NP distribution. The optimal sample temperature is found to depend on the NP diameter. For the 10, 14, and 26 nm NPs, sample temperatures of 475, 500, and 520ºC were found to yield the highest density of NWs, respectively (Table 1). The higher growth temperature for larger particles is expected as the larger particles have higher eutectic temperatures. Notably, while relatively high NW growth yields are observed at the optimal temperatures for both 10 and 14 nm NPs (5-50 NW/$\mu m^2$), significantly lower yield is observed for the 26 nm NPs (~1 NW/$\mu m^2$). This diameter dependence growth yield can be explained by the higher activation energy and higher InAs source delivery rate required for the successful nucleation and therefore growth of larger diameter nanowires. The successful catalytic growth of InAs NWs from Ni NPs arises from the phase properties and the eutectic temperature of the Ni-In-As system. However, the phase diagrams for Ni-In-As have not been well-studied, either experimentally or theoretically. Given the interest in nanowire growth by the vapor-liquid-solid process, this is certainly an area that needs further attention and future exploration. We note that our growth temperatures using Ni and Au NPs of similar diameter are approximately the same (Table 1), suggesting similar eutectic temperatures for the Au-In-As and Ni-In-As systems at this scale.

The structure of the InAs NWs was studied by transmission electron microscopy (TEM). Low and high resolution TEM images confirm the crystallinity and low defect density of the NWs as well as the presence of 2-3 nm thick amorphous surface layer. This layer thickness is consistent with the typical native oxide present on the surface of bulk InAs. Most of the NWs have diameters ranging from ~20-40 nm. The diameters are uniform along the nanowire and no "tapering" is observed, confirming the lack of uncontrolled over-coating during the growth process. Figure 3 shows typical HRTEM images of InAs NWs grown at 475 $^{o}$C. No dominant growth axis was observed. Several NWs studied by TEM grew in the [211] direction, with one such NW shown in Figure 3a. NW growth along the [210] direction (Figure 3b) was also observed. Figure 3c shows a NW grown ~7$^{o}$ off the [111] direction. Data obtained from standardless x-ray EDS elemental analysis gave In/As ratios ranging from 1.2 to 1.5, suggesting the composition of the NWs is close to the expected stoichiometry.

To characterize the electrical properties of the Ni-catalyzed InAs NWs, field-effect transistors (FETs) were fabricated (Fig. 4a) by using Ni (~50 nm) source/drain (S/D) metal contacts in a common back gated geometry (50 nm thermal oxide as gate dielectric, and heavily

B doped Si substrate as the gate). The electrical properties of a representative FET consisting of an individual InAs NW as the channel material with diameter (i.e., channel width) $d$~25 nm (NW diameter ~29 NW with ~2 nm native oxide shell) and a channel length of $L$~9.9µm are show in Fig. 4b-c. The transistor shows a minimal hysteresis (Fig 4b) with an ON current $I_{ON}$~12 µA at $V_{DS}$ = 3V and $V_{GS}$=5V, corresponding to a current density of ~0.5 mA/µm as normalized with the nanowire diameter. Notably, the ON current for this long channel device is comparable to that of the state-of-the-art Si MOSFETs ($I_{ON}$~1 mA/µm), even though the channel length is over two orders of magnitude larger. The device also exhibits a respectable $I_{ON}/I_{OFF}>10^3$ (Fig 4b, inset) at $V_{DS}$=0.5V. The low-bias (low $V_{DS}$, linear-triode region), ON-state conductance of this nanowire as normalized by the channel length is $G_{ON}$~60 µS.µm. Notably, the NW transistors exhibit a uniform response in terms of both ON and OFF state conductance with $G_{ON}$=40-120 µS.µm and $G_{ON}/G_{OFF}>10^2$ for over 100 measured devices ($d$=20-30 nm and $L$=2-10 µm).

The high ON current and conductance of the InAs NWs arise from their high electron mobility, and demonstrates the utility of these synthetic materials for high performance electronics. The field-effect mobility of the nanowires was estimated from the transfer characteristics and is depicted in Fig. 4d as a function of the back-gate voltage $V_{GS}$. The mobility was deduced from the low-bias (low $V_{DS}$) transconductance, $dI_{DS}/dV_{GS}$ of the device by using the standard square-law model, $\mu_n = \frac{dI_{DS}}{dV_{GS}} \times \frac{L^2}{C_{ox}} \times \frac{1}{V_{DS}}$ where $C_{ox}$ is the gate capacitance. The capacitance $C_{ox}$ = 0.52 fF was obtained from modeling using the finite element analysis software Finite Element Method Magnetics. From the square law model, a peak electron mobility of $\mu_n$~2,700 cm$^2$/Vs is obtained. Notably, this electron mobility estimation presents the lower boundary limit as no correction was taken into account, for example, for possible contact resistance. Our Ni-catalyzed InAs NW mobility is comparable to the previously reported Au-catalyzed NWs.[2,12-15] Further enhancement of the mobility may be achieved in the future by the passivation of the NW surfaces with large band-gap InP shell as previously demonstrated in ref. 13.

To further investigate the electrical properties and uniformity of the Ni-catalyzed InAs nanowires, we utilized the contact printing approach[8] to controllably transfer and assemble parallel arrays of NWs on $Si/SiO_2$ substrates over large areas with an average pitch of ~0.5 μm. Poly-L-lysine (0.1%w/v in $H_2O$, Sigma-Aldrich) was applied to the receiver substrate prior to the printing process to enhance the nanowire-substrate chemical interactions and yield a higher nanowire density. Octane:mineral oil (2:1 v/v) was used as a lubricant for the printing process to reduce uncontrollable nanowire-nanowire friction and therefore, enable controlled transfer of aligned nanowires. Following nanowire printing, FETs were fabricated based on the printed InAs NW arrays by using Ni (~50 nm) source/drain (S/D) metal contacts in a common back gated geometry (50 nm thermal oxide as gate dielectric). Figure 5a shows an SEM image and device schematic of a printed InAs NW device. The electrical properties of a representative FET consisting of an array of printed InAs NWs (width~200 μm and channel length~3 μm) are shown in Figures 5b and c. The transistor delivers an ON current of ~6 mA at $V_{DS}$ = 3V, which corresponds to ~15 μA per nanowire (~400 NWs bridging S/D) with $I_{ON}/I_{OFF}$ ~100. The parallel-array NW FETs demonstrate the feasibility of using a printing technology for attaining high performance devices with potentially high switching speeds. Notably, since during the printing process, all synthesized materials are transferred from the growth substrate to the receiver substrate, the results attest the high purity and uniformity of the InAs nanowires grown by using Ni NPs. In future, the nanowire device performance can be readily enhanced through channel length scaling and integration of high-κ dielectrics in a top gate configuration.[13,20-22]

In conclusion, a method to grow crystalline, high-mobility InAs nanowires on amorphous substrates by the VLS/VSS process (non-epitaxial) with the use of Ni catalyst has been demonstraed. Ni NPs are found to serve as efficient catalytic materials for InAs NW growth. Importantly, the chemical composition of the wires grown using this method is close to the expected stoichiometry, therefore, enabling high performance nanowire devices with electron mobility ~2,700 $cm^2/Vs$ and $I_{ON}/I_{OFF}>10^3$. The high yield and purity of the grown InAs nanowires enable for the successful and aligned transfer of NWs from the growth substrate to the receiver substrate by the contact printing process. The ability to grow high mobility, crystalline, CMOS-compatible, and printable InAs NWs may have important implications for future integration of InAs NWs for various electronic applications.

## Experimental Section

Thermal evaporation was carried out using 99.995% pure Ni pellets (Kurt J. Lesker) under a vacuum of ~$8x10^{-7}$ torr. AFM used to image the Ni catalysts after annealing was performed using a Digital Instruments Dimension 3100. SEM and TEM were performed on the InAs nanowires using a Gemini Leo 1550 and JEOL 2100-F 200 kV, respectively. The InAs nanowire FET devices were fabricated by drop-casting the InAs nanowires suspended in ethanol onto p-type Si substrates with a 50 nm gate oxide. Photolithography was used to define the source and drain regions and Ni was thermally evaporated to form the source and drain contacts. Details of the printing process for fabrication of the InAs NW array FETs have been previously described.[8]

**Acknowledgements.** This work was financially supported by MARCO/MSD Focus Center Research Program, Intel Corporation, Lawrence Berkeley National Laboratory, and an Intel

Graduate Fellowship (J.C.H.). All fabrication was performed in the UC Berkeley Microlab facility. TEM imaging was performed at the Molecular Foundry, Lawrence Berkeley National Laboratory, which is supported by the Office of Science, Office of Basic Energy Sciences, U.S. Department of Energy, under Contract No. DE-AC02-05CH11231.

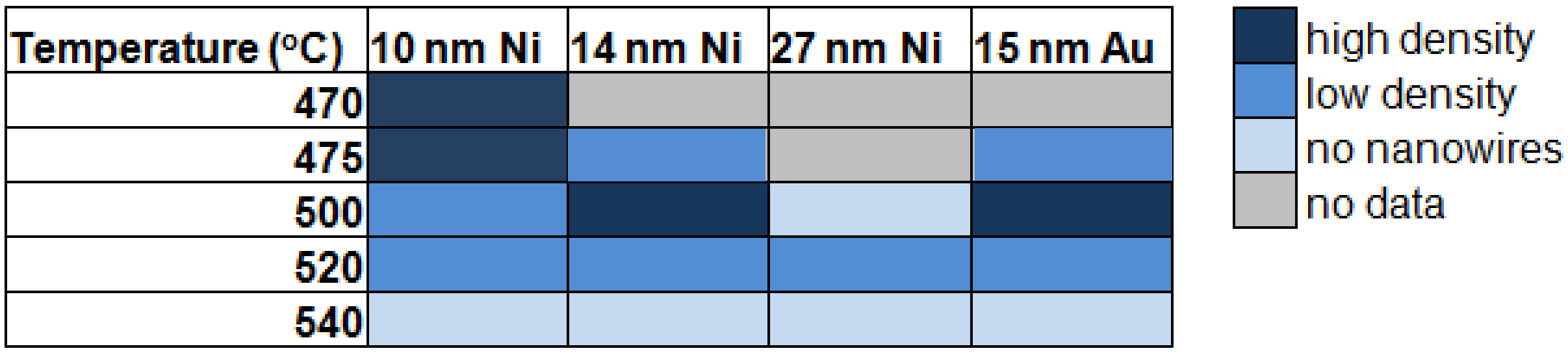

**Table 1.** InAs NW growth yield studies at various sample temperatures for annealed Ni films of various thicknesses and commercially available Au colloids. High density corresponds to >5 NW/µm$^2$ while low density corresponds to ~1 NW/µm$^2$.

## Figure Captions

**Figure 1**. AFM images and NP diameter distribution histograms for Ni particles resulting from the thermal anneal of (a,b) ~0.5 nm, (c,d) ~1.5 nm, and (e,f) ~3 nm Ni films at 850 ºC for 10 min. All AFM images show an area of 1 μm x 1 μm. Particle diameter average and standard deviation are shown in the upper right corners of the histograms.

**Figure 2**. SEM images and nanowire diameter distribution histograms for InAs nanowires grown using Ni catalyst particles produced by the thermal anneal of (a,b) 0.5 nm, (c,d) 1.5 nm, and (e,f) 3 nm Ni films. SEM image insets clearly show the Ni catalyst tips at the ends of the nanowires, depicting the tip-based growth mechanism. Nanowire diameter average and standard deviation are shown in the upper right corners of the histograms.

**Figure 3**. HRTEM images of typical InAs NWs grown using Ni catalyst NPs. (a) Growth axis along the [211] direction, (b) growth axis along the [210] direction, and (c) growth axis $7^{o}$ off the [111] direction.

**Figure 4**. (a) A SEM image and a schematic of a back-gated InAs nanowire FET with Ni S/D metal contacts. (b) Linear scale, transfer characteristics of a representative Ni-catalyzed InAs FET with $d$~29 nm (~25 nm InAs core with ~2 nm thick native oxide shell) and $L$~9.9 for $V_{DS}$ = 0.1, 0.3, and 0.5 V. Both forward and backward gate voltage sweep directions are shown, exhibiting a minimal hysteresis. The inset shows the log scale $I_{DS}$-$V_{GS}$ curve for $V_{DS}$ = 0.5 V. (c) Output characteristics of the same nanowire device at various $V_{GS}$. (d) Electron mobility vs. $V_{GS}$ estimated from the transfer characteristic at $V_{DS}$ = 0.1 V using the square-law model. The black

dotted line shows the actual data with a peak mobility of ~3000 $cm^2/Vs$ and the solid red line shows the smoothed values (peak mobility ~2700 $cm^2/Vs$). All electrical measurements were conducted in vacuum in order to minimize the hysteresis and noise due to the ambient environment.

**Figure 5.** Contact printed NW devices. (a) An SEM image and a device schematic of a back-gated FET fabricated on a printed, parallel array of InAs nanowires. (b) Linear scale, transfer characteristics of a representative FET with *W*~200 μm (~400 NWs bridging S/D) and *L*~3 μm at $V_{DS}$ = 0.1, 0.3, and 0.5 V. The inset shows the log scale $I_{DS}$-$V_{GS}$ curve for $V_{DS}$ = 0.3 V. (c) Output characteristics of the same device at various $V_{GS}$.

# Figure 1

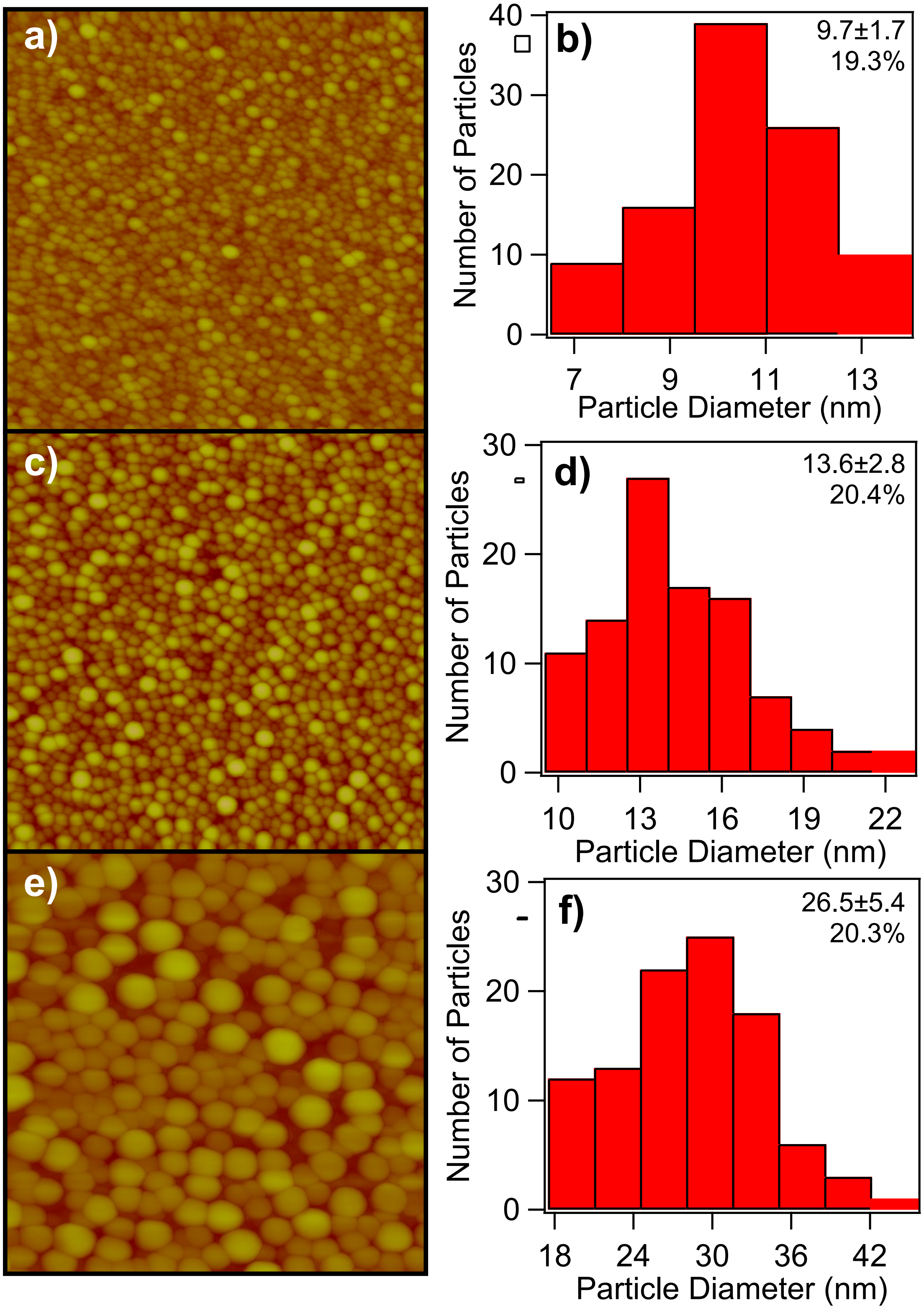

a)
b)
9.7±1.7
19.3%
Number of Particles
Particle Diameter (nm)
c)
d)
13.6±2.8
20.4%
Number of Particles
Particle Diameter (nm)
e)
f)
26.5±5.4
20.3%
Number of Particles
Particle Diameter (nm)

# Figure 2

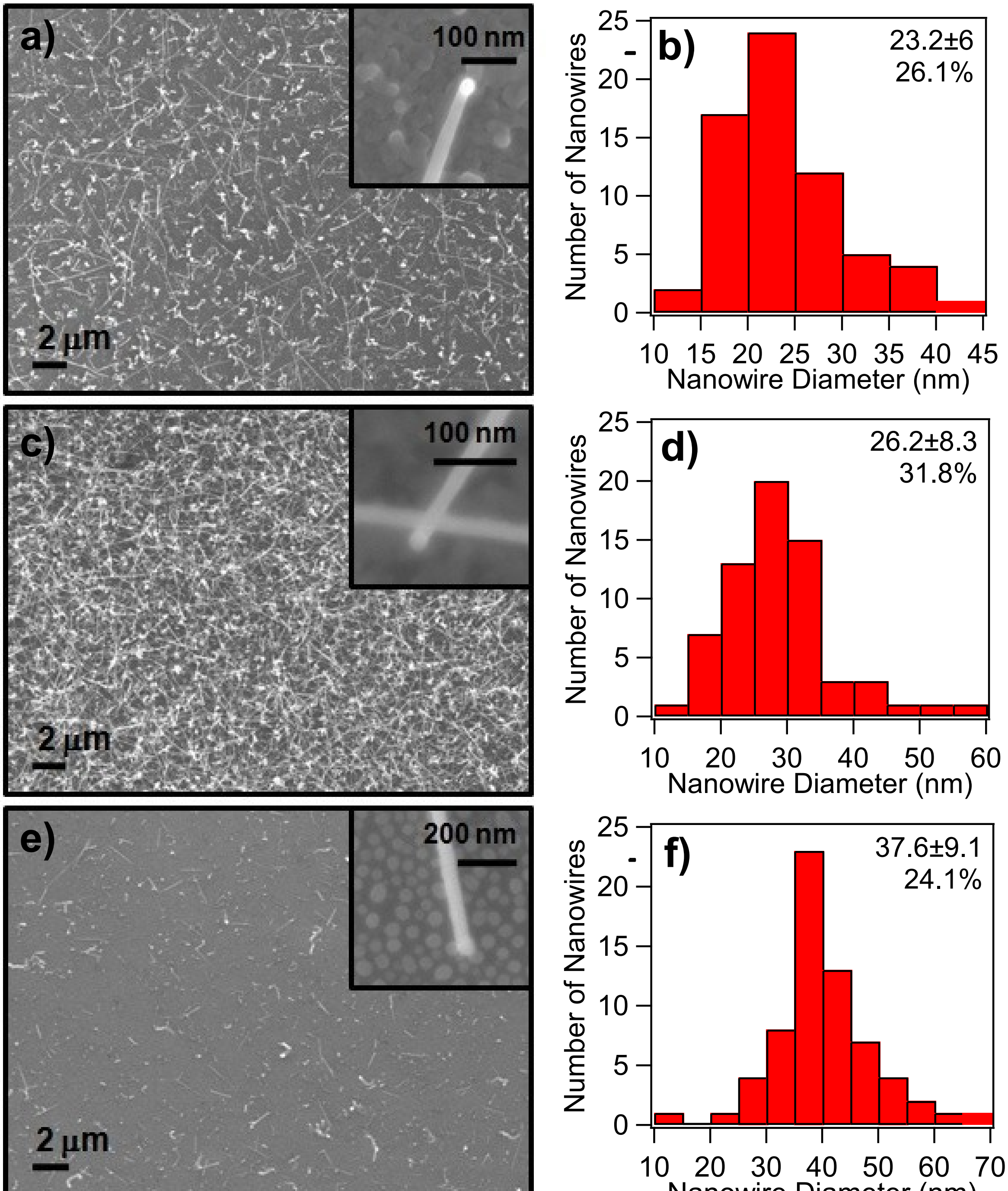

a)
100 nm
2 µm
b)
23.2±6
26.1%
Number of Nanowires
Nanowire Diameter (nm)
c)
100 nm
2 µm
d)
26.2±8.3
31.8%
Number of Nanowires
Nanowire Diameter (nm)
e)
200 nm
2 µm
f)
37.6±9.1
24.1%
Number of Nanowires
Nanowire Diameter (nm)

# Figure 3

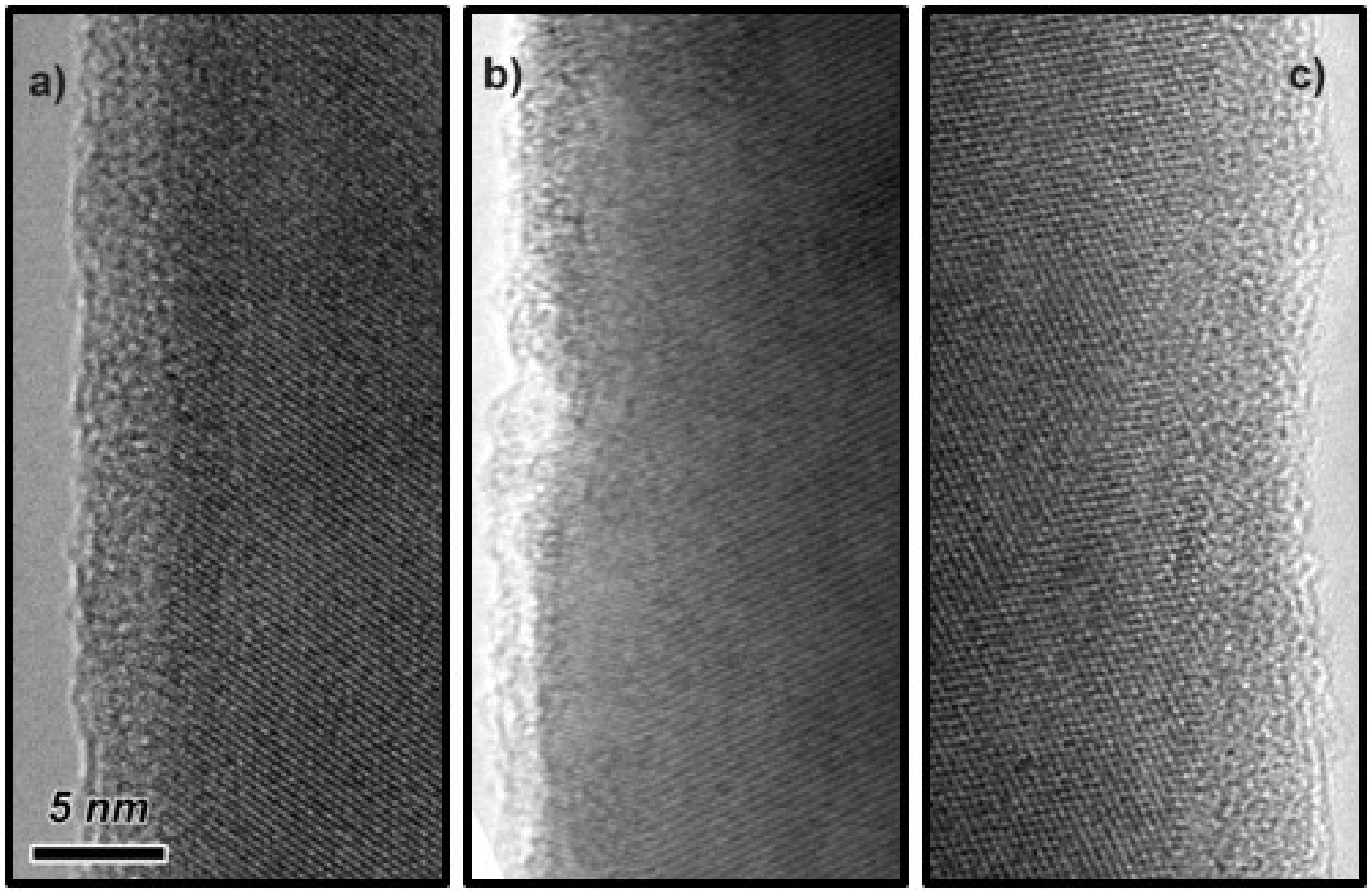

# Figure 4

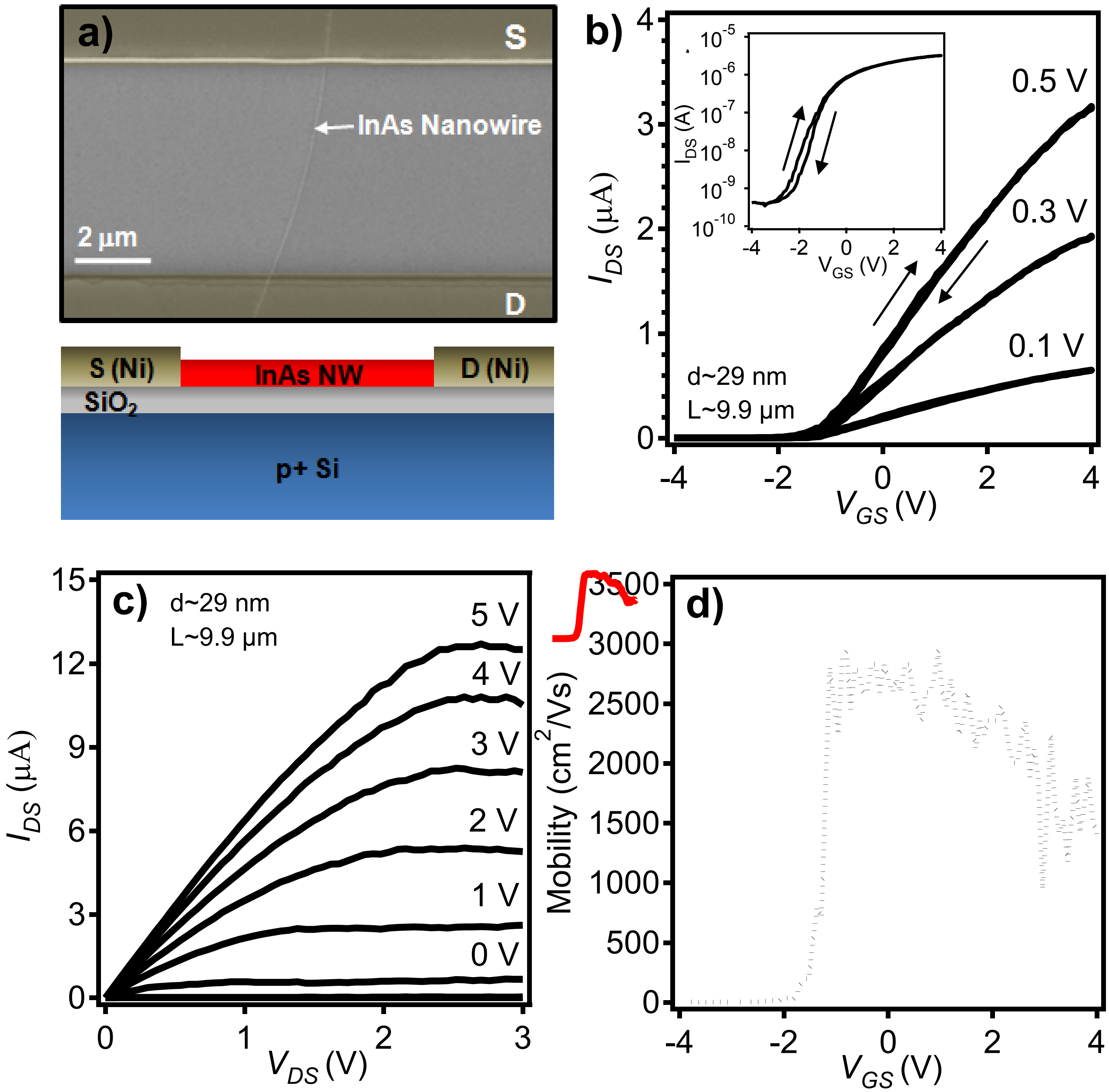

a)
S
InAs Nanowire
2 µm
D
S (Ni)
InAs NW
D (Ni)
SiO2
p+ Si
b)
0.5 V
0.3 V
0.1 V
d~29 nm
L~9.9 µm
IDS (µA)
VGS (V)
IDS (A)
c)
d~29 nm
L~9.9 µm
5 V
4 V
3 V
2 V
1 V
0 V
IDS (µA)
VDS (V)
d)
Mobility (cm2/Vs)
VGS (V)

# Figure 5

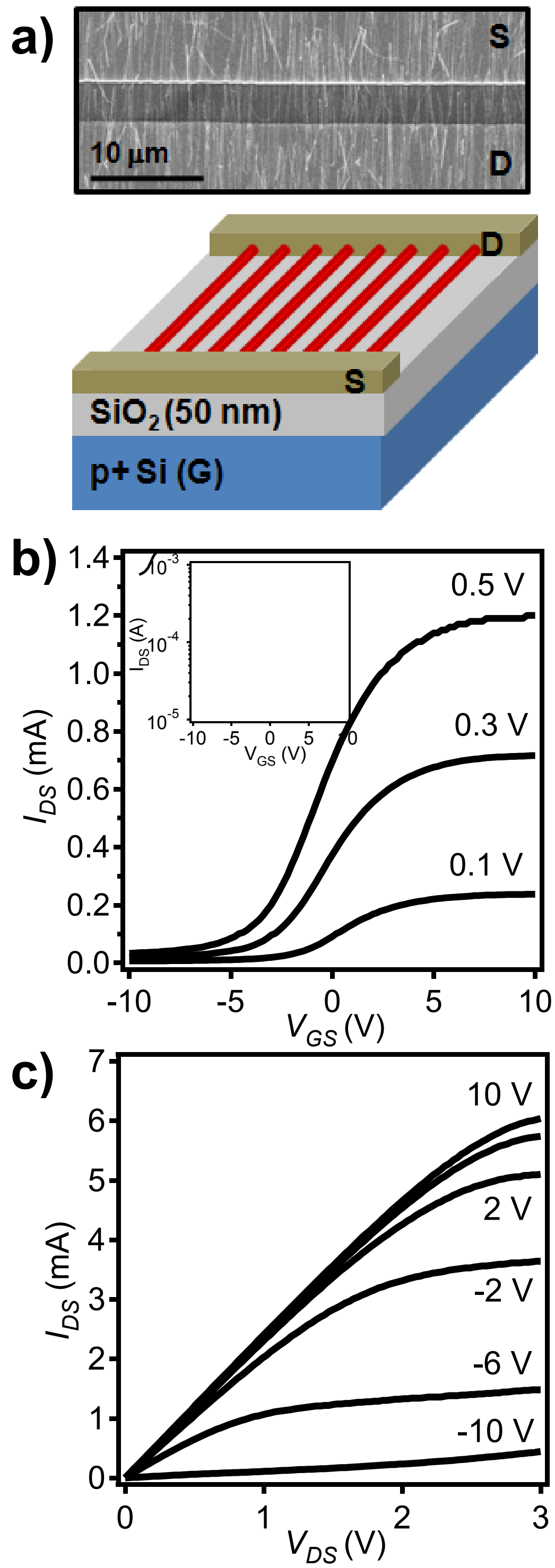

a)
S
10 µm
D
D
S
SiO2(50 nm)
p+ Si (G)
b)
0.5 V
0.3 V
0.1 V
IDS (mA)
VGS (V)
IDS (A)
VGS (V)
c)
10 V
2 V
-2 V
-6 V
-10 V
IDS (mA)
VDS (V)

## Table of Contents

Ni nanoparticles are shown to serve as an efficient catalyst for high yield growth of high mobility InAs nanowires on amorphous $SiO_2$ substrates. The grown nanowires can be readily printed as parallel arrays on substrates and configured as high performance transistors.

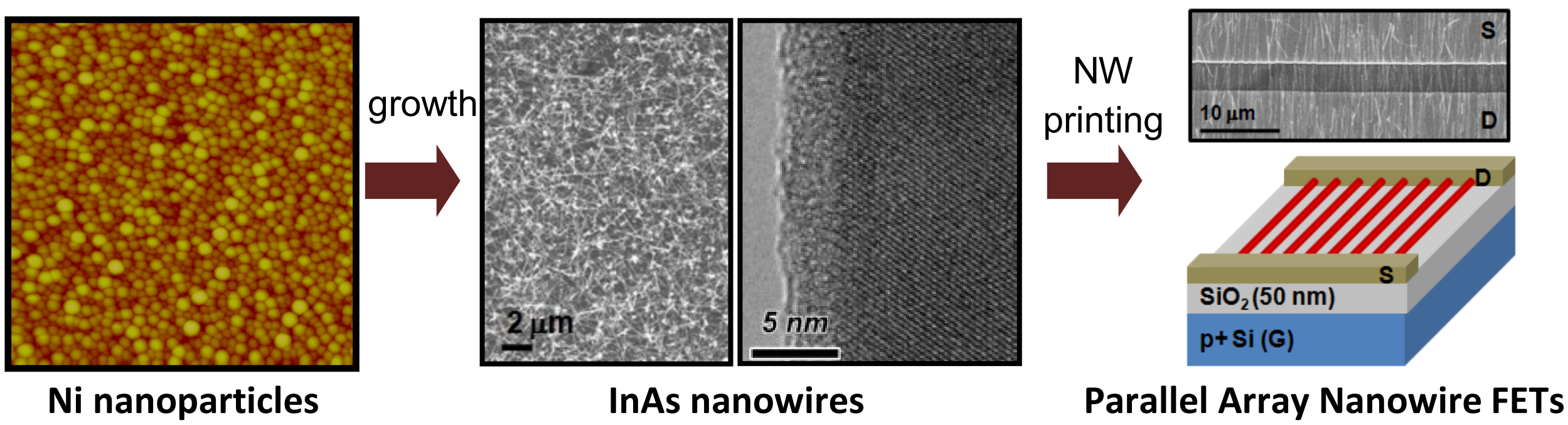